\documentclass[10pt]{IEEEtran}
\usepackage{amsmath}
\usepackage{bm}
\usepackage{amssymb}
\usepackage{latexsym}
\usepackage{multirow}
\usepackage{epsfig}
\usepackage{graphics}
\usepackage{mathrsfs}
\usepackage{algorithm}
\usepackage{algpseudocode}
\usepackage{subcaption}
\usepackage{color}
\usepackage{graphicx}
\usepackage{algcompatible}
\usepackage{array}
\usepackage{amsfonts}
\usepackage{mdwmath}
\usepackage{mdwtab}
\usepackage{eqparbox}
\usepackage{epstopdf}
\newcommand{\bc}{\begin{center}}
\newcommand{\ec}{\end{center}}
\newcommand{\be}{\begin{equation}}
\newcommand{\ee}{\end{equation}}
\newcommand{\bea}{\begin{eqnarray}}
\newcommand{\eea}{\end{eqnarray}}

\newtheorem{theorem}{Theorem}[section]

\newtheorem{lemma}[theorem]{Lemma}

\newcounter{tempEquationCounter}
\newcounter{thisEquationNumber}
\newenvironment{floatEq}
{\setcounter{thisEquationNumber}{\value{equation}}\addtocounter{equation}{1}% record equation as happened and remember number
\begin{figure*}[!t]% float following equation across columns
\normalsize\setcounter{tempEquationCounter}{\value{equation}}% record current equation number in floated location
\setcounter{equation}{\value{thisEquationNumber}}% use previous equation number
}
{\setcounter{equation}{\value{tempEquationCounter}}% set back to equation number in floated location
\hrulefill\vspace*{4pt}% add a horizontal rule separator
\end{figure*}% end float environment

}
\begin{document}

\title{Towards Green Mobile Edge Computing Offloading Systems with Security Enhancement}
 \author{Haijian Sun, \emph{Member, IEEE}, Qun Wang, \emph{Student Member, IEEE}, Xiang Ma, \emph{Student Member, IEEE}，\\ Yongjun Xu, \emph{Member, IEEE}, and Rose Qingyang Hu, \emph{Fellow, IEEE}\\

\thanks{H. Sun is with the Department of Computer Science, University of Wisconsin-Whitewater, Whitewater, WI, USA. Q. Wang, X. Ma, and R. Q. Hu is with Electrical and Computer Engineering Department, Utah State University, Logan, UT, USA. Y. Xu is with with the School of Communication and Information Engineering,
Chongqing University of Posts and Telecommunications, Chongqing 400065,
China. Y. Xu is also with the Shandong Provincial Key Lab of Wireless
Communication Technologies, Shandong University, Jinan 250100, China. } }

\maketitle
\begin{abstract}
Mobile edge computing (MEC) is an emerging communication scheme that aims at reducing latency. In this paper, we investigate a green MEC system  under the existence of an eavesdropper. We use computation efficiency, which is defined as the total number of bits computed divided by the total energy consumption, as our main metric. To alleviate stringent latency requirement, a joint secure offloading and computation model is proposed. Additionally, we formulate an optimization problem for maximizing computation efficiency, under several practical constraints. The non-convex problem is tackled by successive convex approximation and an iterative algorithm. Simulation results have verified the superiority of our proposed scheme, as well as the effectiveness of our problem solution. 
\end{abstract}

\section{Introduction}
To meet the ever increasing latency requirement over wireless connections, mobile edge computing (MEC) has emerged to be a promising scheme for the next generation wireless system. Compared with cloud computing, MEC utilizes computation resources at the network edge hence transmission to super data center farther away can be largely avoided \cite{haijian}. 

One of the main directions for MEC research is to exploit the advantage for data processing, or computation. Specifically, if a user (UE) has a large amount of data to be computed with time budget, it can send part of the data to MEC server and leave another part locally for simultaneous processing. After calculation,  the server then sends the computed result back to users. This so-called joint offloading and computing scheme has received extensive attentions recently \cite{Sbi}-\cite{LYang}. In general, there are two categories in offloading, namely 1) the binary model, which allows the user to either offload all computing bits to server (similar to cloud setting),  2) the joint model, which dynamically partition task into two parts: offloading and local computing. In this paper, we consider the latter case. 

Different from existing works that focus on either computation data maximization or energy consumption minimization, our prior work \cite{meMEC} proposed computation efficiency, similar to energy efficiency in wireless communication, is defined as the number of bit computed divided by the total energy consumed. Several subsequent papers have adopted our metric and studied its performance under different scenarios. For example, \cite{fuhui1} investigated computation efficiency maximization under wireless-powered MEC systems. It also compared two offloading schemes: joint and binary.  An orthogonal frequency division multiple access (OFDMA)-based MEC system with computation efficiency maximization is proposed in \cite{fuhui2}, aiming for power and channel optimal allocation. Lastly, \cite{fuhui3} applied computation efficiency in an unmanned aerial vehicle (UAV)-enabled MEC offloading system, which also seeks for computation efficiency maximization.  

On the other hand, wireless offloading will unavoidably suffer from potential malicious activities due to the existence of eavesdropper.  From physical layer information-theoretic perspective, achievable data rate with eavesdropper can be modeled as the difference of mutual information from transmitter to receiver and from transmitter to eavesdropper, regardless of security protection mechanisms. It can be considered as the lower bound of the actual data rate. This simplified analysis model has been adopted in various papers \cite{phy1}-\cite{phy2}.

Our contributions are summarized as follows. 
\begin{enumerate}
\item We consider a secure offloading model, which allows wireless transmission with the presence of eavesdroppers. We adopt the physical layer security model from information-theoretic perspective and is irrelevant of  encryption schemes. 

\item Computation efficiency is applied as the main metric, which finds the balance between maximizing computation bits and minimizing total energy consumption. 

\item An iterative algorithm together with convex approximation is proposed to tackle a non-convex problem and has good convergence speed and performance. 
\end{enumerate}

This paper is organized as follows. Section II describes the secure offloading and computation model. An optimization problem is formulated in Section III, followed by its approximation solution. Numeric results are presented in Section IV.  Finally, Section V concludes this paper.

\section{System Model}
We consider a typical MEC system with one server and $K$ UEs, the MEC system also mounts a wireless access point (AP) to communicate with other devices. We assume that both the AP and the UEs have a single antenna. There also exists a malicious eavesdropper that tries to intercept confidential information. Denoted as $Eve$, the eavesdropper only has one antenna. At the beginning of a reference time $t_s$, each UE has a large number of computation-intensive task to be computed, because of the limited computation resources at UE, due to either device size or power constraints or both, UEs cannot finish their tasks before $t_e$. For timely processing, we require $t_e - t_s \leq T$. Hence, UEs will offload part of their computation bits to the MEC server, where a more powerful processing can be supported. In general, each UE supports the following operation modes.
\begin{figure}[h]
  \centering
  \includegraphics[width=2.5in]{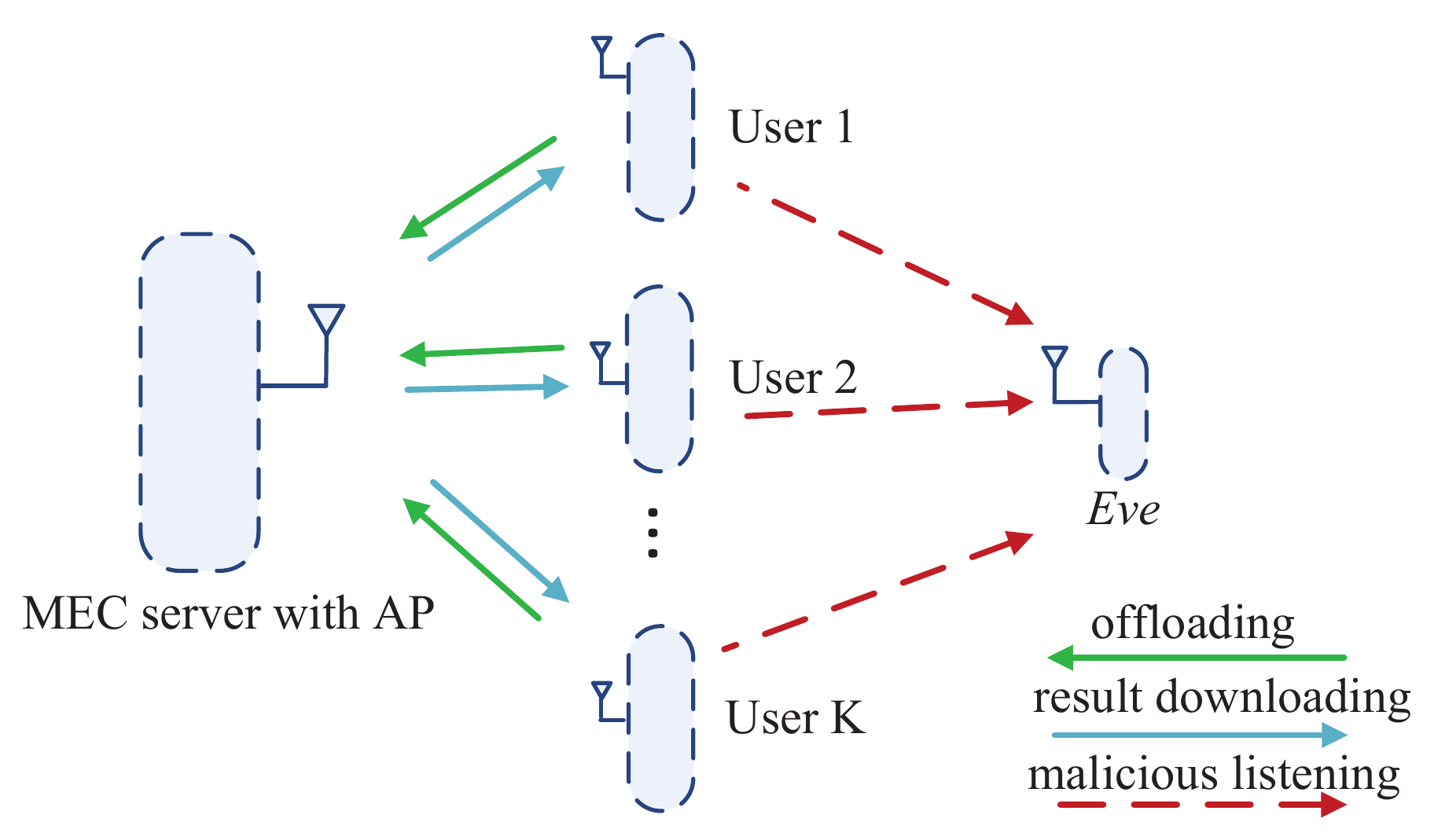}\\
  \caption{Secure MEC partial offloading model}\label{1}
\end{figure}
\subsection{Secure Offloading}
In the presence of $Eve$, each UE must securely offload part of their task to the MEC server.  We assume the channel between each UE and the AP in MEC server follows block static model where it remains the same within the block time $T_1$ ($T_1 \geq T $) but varies from one block to another. We denote the channel between UE $k$ and the AP be $h_k$, where $h_k = l_k  h_0$ is the joint effect of large-scale $ l_k$ and small-scale $h_0$ fading. Similarly, the channel between each UE and the $Eve$ is $g_k$.

We consider the \emph{active} eavesdropper scenario, where the $Eve$ is also a user in the system, its listening and transmitting can be captured by UEs (from authentication, etc). Therefore in this setting, we assume the channel between UE $k$ and the $Eve$ can be perfectly estimated, i.e., $g_k$ can be perfectly estimated. Similar setting can also be found in \cite{phy2}.

Let the number of total bits be computed for each UE be $L_k$, since each UE cannot finish the calculation before the required time slot, it will send to MEC server for joint processing. Specifically, $m_k$ is the number of bits that UEs offloads to the MEC securely. The signal received at the AP and $Eve$ become: 

\bea
y_k = h_k^H s_k + n_k, \forall k=1, \ldots, K,  \\
y_e^k = g_k^H s_k + n_e^k, \forall k=1, \ldots, K,
\eea
here, $s_k \in \mathbb{C}$ is the information-bearing signal for UE $k$,  $n_k \in \mathcal{CN} (0, \sigma_k^2)$ and ${n}_e^k \in \mathcal{CN} (0, \sigma_{ek}^2)$ are the complex Gaussian noise at the AP and $Eve$, respectively. The secrecy rate, from information-theoretic perspective, is given as 
\be
R_{k,a}^{\text{sec}} =  \big[  \log ( 1 + \frac{p_k h_k^2  }{\sigma_k^2} )    - \log \big(1+\frac{ p_k  g_k^2} {\sigma_{ek}^2}  \big)\big]^+,
\ee
where $[a]^+ = \max(a, 0)$.

For offloading, the energy consumption consists of two parts: transmission and fixed circuit. In particular, $E_k^{\text{off}} = p_k t_k + p_r t_k$, where $p_r$, a constant, is the power of other circuit except for the transmission unit. 

The rest $L_k - m_k$ bits will be calculated locally, which will be described below. 

\subsection{Local Computing}
Traditionally, users will process all the computation locally. To model such a process, we first define some parameters. First, we assume user $k$'s CPU needs $C_k$ cycles to finish the computation of a single bit of data. Also, let $f_k$ be the clock speed of the CPU. For simplicity, we assume the clock speed does not change. Each user is allowed to start the local computing from the beginning to the end of the process, thus the total number of computation bits becomes $T f_k / C_k$.   

Energy consumption for local computing can be modeled as $E_k^{\text{comp}} = \epsilon_k f_k^3 T$, where $\epsilon_k$ is the CPU energy coefficient \cite{QWu}, \cite{YWang}. 

\subsection{Receiving Computed Results}
After receiving the computation task from each user, MEC server will start the calculation. When finished, it will send the result back to each user. Here, like \cite{LYang} \cite{fuhui1}, we assume this process takes negligible time because of two reasons: 1) MEC server has powerful multi-thread processor, 2) compared with data bits to be computed, result takes way less space, hence downlink transmission is almost instant.  

\begin{figure}[h]
  \centering
  \includegraphics[width=3in]{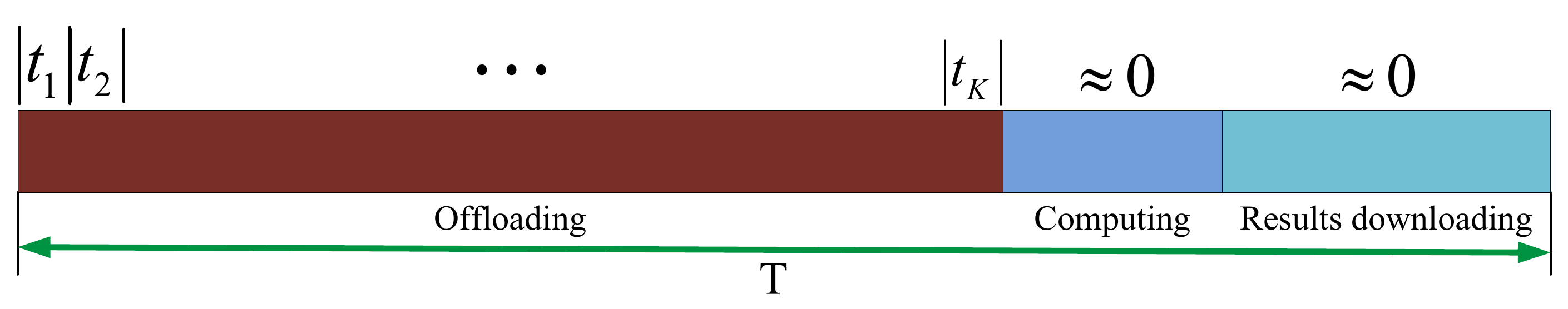}\\
  \caption{Time sharing offloading scheduling}\label{fig2}
\end{figure}

The whole process is illustrated in Fig. \ref{fig2}.

\subsection{Computation Efficiency in MEC Systems}
Like previously mentioned, we define computation efficiency as the total number of calculated bits divided by total energy consumed. Thus, $CE_k =\frac{B R_{k,a}^{\text{sec}} t_k + \frac{T f_k}{ C_k} }{p_k t_k + p_r t_k + \epsilon_k f_k^3 T}$ \cite{meMEC}, where $B$ is the bandwidth for offloading. 
\section{Computation Efficiency Maximization with Active Eavesdropper}

We consider a green MEC system where the objective is to maximize the computation efficiency, the optimization problem is formulated as follows.
\begin{subequations}
\begin{eqnarray}
\mathbf{P}_1 & &  \kern-1em \kern-1em \kern-1em \max_{ \substack{ \{t_k\}, \{f_k\}, \\ \{m_k\}, \{p_{k}\}} } \ \sum_k w_k \frac{B R_{k,a}^{\text{sec}} t_k  + \frac{T f_k}{C_k}}  {\epsilon_k f_k^3 T + p_k t_k + p_r t_k}     \\
\text{s.t.} \   C1 &:& \sum_k t_k \leq T, \label{p1c1}\\
C2 &:& B R_{k,a}^{\text{sec}} t_k \geq m_k, \forall k, \label{p1c2}\\
C3 &:&  L_k - \frac{T f_k^{\text{max}}}{C_k} \leq m_k \leq L_k, \forall k, \label{p1c3} \\
C4 &:& \epsilon_k f_k^3 T +  p_k t_k + p_r t_k \leq E_{k}^{th}, \forall k, \label{p1c4}\\
C6 &:& 0 \leq f_k \leq f_k^{\text{max}}, t_k \geq 0 , p_k \geq 0 \ \forall k \label{p1c6}
\end{eqnarray}
\end{subequations}

Our objective is to find the maximum value of the weighted summation for each UE's computation efficiency, $w_k$ is the weight for UE $k$. The variables to be optimized here is the transmitted time for each UE $t_k$, the CPU frequency $f_k$, and the transmitted power for each user $p_k$. (\ref{p1c1}) is the time constraint which requires the whole process ends before time $T$, (\ref{p1c2}) combined with (\ref{p1c3}) is the requirement for offloading and local computing rate. Furthermore, (\ref{p1c4}) is the energy consumption constraint for each UE, where $E_{k}^{th}$ is the maximum allowed energy.  Lastly, $f_k$, $t_k$, and $p_k$ should be non-negative variables, which is defined in (\ref{p1c6}) \footnote{Theoretically, $m_k$ should be an integer value, and (\ref{p1c3}) becomes $m_k \in \{ L_k - \frac{T f_k^{\text{max}}}{C_k}, L_k - \frac{T f_k^{\text{max}}}{C_k} + 1, \ldots, L_k\}$. However, we simplify this and allow $m_k$ to be fractional value, a feasible solution can take the round value if it is fractional, in \cite{Sbi}, they also take the same approximation. When $L_k$ is very large as the most real-world cases, the approximation has minimal impact to the original problem.}.

Clearly, the formulated problem is non-convex, due to its sum-of-ratio objective function and \emph{C2, C4, C5}, especially the coupling variables $p_k$  and $t_k$.  In the following, we tackle each non-convex term, mainly with successive convex approximation (SCA).

\subsection{SCA-based Optimization Algorithm}

Firstly, we transform (\ref{p1c2}) and (\ref{p1c4}). Notice that we group these two constraints since they both involve with coupling variables. Let $\tilde{p}_k  = {p}_k t_k$,  then $ {R}_{k,a}^{\text{sec}}  t_k = t_k  \log \big(1  + \frac{1} {\sigma_k^2}   \tilde{p}_k h_k^2 / t_k \big)  -  t_k \log \big(1+  \frac{1} {\sigma_{ek}^2}  \tilde{p}_k g_k^2  / t_k \big)$. For a function in the form of $f(x,y) = y\log(1+\frac{x}{y})$, it represents the entropy between $x$ and $y$, and it is a concave function. Thus, $ {R}_{k,a}^{\text{sec}} $ is still non-convex due to the difference of concave and convex functions. For approximation, we apply SCA algorithm. Specifically, the entropy function has first-order Taylor series expansion at $(x,y) = (x_0,y_0)$  
\bea  \nonumber
f(x,y) = y_0 \log (1+ \frac{x_0}{y_0}) +  [\log(1+\frac{x_0}{y_0}) - \frac{x_0}{x_0+y_0}] (y-y_0) \\ 
 +    \frac{y_0}{x_0+y_0} (x-x_0),  \nonumber
\eea
where  $(x_0,y_0)$ is the differentiable point. It is easy to verify that given  $(x_0,y_0)$,  $f(x,y)$ becomes an affine form, which is convex. In optimization problems, we solve for $ (x,y)$ with given feasible point  $(x_0,y_0)$ first, then in the next round,  $(x_0,y_0)$ becomes the previous round's $ (x,y)$. The process will continue until converges.  SCA-based approach works well in the iterative algorithms and received much attention recently.

In the following, to simplify notations, we first transform (\ref{p1c3}) to equation sets below: 
\begin{subequations} \label{transC2}
\begin{eqnarray} \label{transC21}
\tau_k \geq  \frac{1} {\sigma_{ek}^2}  { \tilde{p}_k g_k^2},  \label{form1}\\ 
t_k  \log \big( 1 + \frac{N_k} {t_k} \big) - t_k \log \big( 1+ \frac{\tau_k}{t_k} \big) \geq \frac{m_k}{B}, \label{taylor1}\\ \label{transC22}
N_k \leq  \frac{1} {\sigma_k^2}    \tilde{p}_k h_k^2, \label{4c} \label{transC23}
\end{eqnarray}
\end{subequations}
where $\tau_k$ and $N_k$ are auxiliary variables.
It is easy to verify that the transformation is equivalent. 
 Furthermore, (\ref{taylor1}) should be transformed according to $f(\tau_k,t_k)$ and $f(N_k,t_k)$ that mentioned above.

\subsection{Objective Function}
Next, the objective function needs to be converted to the convex form as well. Currently, it is the summation of the fractional functions (represent each UE's computation efficiency).  Traditional Dinkelbach's method cannot be applied directly since it can only deal with one fractional function. Instead, following \cite{YJong}, we can generalize Dinkelbach's algorithm to tackle one fractional function to multiple ones by a simple transformation.

\begin{subequations}
\begin{eqnarray}
\mathbf{P}_3 & &  \kern-1em \kern-1em \kern-1em \max_{ \{t_k\}, \{f_k\},\{m_k\}, \{ \tilde{p}_k \},\{\beta_k\}, \{\tau_k\}, \{ N_k \} } \ \sum_k w_k \beta_k\\
\text{s.t.} &&  C1 : R_k \geq \beta_k  E_k, \forall k,   \\
C2 &:& \sum_k t_k \leq T, \label{p3c1}\\
C3 &:&  L_k - \frac{T f_k^{\text{max}}}{C_k} \leq m_k \leq L_k, \forall k, \label{p3c3} \\
C4 &:& \epsilon_k f_k^3 T + \tilde{p}_k + p_r t_k \leq E_{k}^{th}, \forall k, \label{p3c4}\\
C5 &:& 0 \leq f_k \leq f_k^{\text{max}}, t_k \geq 0 , \tilde{p}_k \geq  0 \ \forall k, \label{p3c6} \\
C6 &:& (\ref{transC21})  - (\ref{transC23}). \label{p3c7}
\end{eqnarray}
\end{subequations}
Here, we let $R_k = B R_k^{\text{sec}} t_k  + \frac{T f_k}{C_k}$, and $E_k =  {\epsilon_k f_k^3 T +   \tilde{p}_k + p_r t_k}$, for notational simplicity. $\beta_k$ is the auxiliary variable. 

\begin{floatEq}
\begin{subequations}
\begin{eqnarray}
 & &     \kern-1em   \kern-1em \kern-1em \kern-1em \kern-1em \max_{\substack{ \{t_k\}, \{f_k\},   \{m_k\}, \\ \{ \tilde{p}_k \}, \{\tau_k\}, \{ N_k \} }}   \sum_k \lambda_k w_k B  \bigg \{( v_k^{(i)}-\theta_k^{(i)}) (t_k - t_k^{(i)})+ \frac{t_k^{(i)}}{t_k^{(i)} + N_k^{(i)}} (N_k - N_k^{(i)})  - \frac{t_k^{(i)}}{t_k^{(i)} + \tau_k^{(i)}} (\tau_k - \tau_k^{(i)}) \bigg\} + \sum_k \frac{ \lambda_k w_k  T f_k}{C_k}  \\ \nonumber  && -\sum_k  \lambda_k \beta_k \big(\epsilon_k f_k^3 T +   \tilde{p}_k + p_r t_k \big) \\
& & \text{s.t.}  \  B \bigg \{( v_k^{(i)}-\theta_k^{(i)}) (t_k - t_k^{(i)})+\frac{t_k^{(i)}}{t_k^{(i)} + N_k^{(i)}} (N_k - N_k^{(i)})  - \frac{t_k^{(i)}}{t_k^{(i)} + \tau_k^{(i)}} (\tau_k - \tau_k^{(i)})+ \varphi_k^{(i)} \bigg\}  + \frac{ T f_k}{C_k} \geq L_k  , \\
& & (\ref{p3c1}), (\ref{p3c4}), (\ref{p3c6}), (\ref{transC21}), (\ref{transC23})
\label{P4}
\end{eqnarray}
\end{subequations}
\end{floatEq}

$\mathbf{P}_3$ can be solved with the following \emph{Lemma}.
\begin{lemma}
For $\forall k$, if $ (\{t_k^*\}, \{f_k^*\},\{m_k^*\}, \{ \tilde{p}_k^* \},\{\beta_k^*\}, \{\tau_k^*\}, \{ N_k^* \} )$ is the optimal solution of $\mathbf{P}_3$, there must exist $\{\lambda_k^*\}$ such that $ (\{t_k^*\}, \{f_k^*\},\{m_k^*\}, \{ \tilde{p}_k^* \}, \{\tau_k^*\}, \{ N_k^* \} )$  satisfies the Karush-Kuhn-Tucker (KKT) condition of the following problem for $\lambda_k = \lambda_k^*$ and $\beta_k = \beta_k^*$.
\begin{subequations}
\begin{eqnarray}
 \mathbf{P}_4 :  & &  \max_{ \{t_k\}, \{f_k\}, \{P_{k}\}} \sum_k  \lambda_k (w_k R_k  - \beta_k E_k )   \label{p4_obj} \\
\text{s.t.} & & (\ref{p3c1}) - (\ref{p3c7}).
\end{eqnarray}
\end{subequations}
Furthermore,  $ (\{t_k^*\}, \{f_k^*\},\{m_k^*\}, \{ \tilde{p}_k^* \}, \{\tau_k^*\}, \{ N_k^* \} )$  satisfies the following equations for $\lambda_k = \lambda_k^\star$ and $\beta_k = \beta_k^\star$:
\be
\lambda_k = \frac{w_k}{E_k }, \ \beta_k = \frac{w_k R_k }{ E_k }, \forall k. \label{KKT}
\ee
\end{lemma}
Please refer to \cite{YJong} for the detailed proof. 

Based on the above \emph{Lemma}, $\mathbf{P}_3$ can be solved iteratively. At each iteration, the objective function becomes a convex one with giving $\lambda_k$ and $\beta_k$ in (\ref{p4_obj}), then the auxiliary value of $\lambda_k$ and $\beta_k$ will be updated according to the next section.

\subsection{Proposed Solution to $\mathbf{P}_4$ with given $(\lambda_k, \beta_k)$}

To summarize, for given $(\lambda_k, \beta_k)$, a complete version of  $\mathbf{P}_4$ is illustrated at the top of next page. Where $\theta_k^{(i)} =  [\log(1+\frac{\tau_k^{(i)}}{t_k^{(i)}}) - \frac{\tau_k^{(i)}}{\tau_k^{(i)}+t_k^{(i)}}]$, $ v_k^{(i)} =  [\log(1+\frac{N_k^{(i)}}{t_k^{(i)}}) - \frac{N_k^{(i)}}{N_k^{(i)}+t_k^{(i)}}]$, and $\varphi_k^{(i)} = t_k^{(i)}   \log ( 1 + N_k^{(i)}  /  t_k^{(i)} ) -  t_k^{(i)}  \log(1+   \tau_k^{(i)}  / t_k^{(i)} ) $ is replaced for notational simplicity.

It is easy to verify that, for given $ v_k^{(i)}, \theta_k^{(i)}, t_k^{(i)}, N_k^{(i)}$, and $\tau_k^{(i)}$ at each iteration, $\mathbf{P}_4$ is a convex optimization problem and can be solved by standard method such as interior-point algorithm. The optimized variable, once computed from the optimization problem, will be used to update $ v_k^{(i)}, \theta_k^{(i)}, t_k^{(i)}, N_k^{(i)}$, and $\tau_k^{(i)}$ for the input of next iteration. Furthermore, the convergence of this iterative algorithm can be guaranteed by concave-convex procedure (CCCP).

Notice that for SCA algorithm, it is vital to select an appropriate initial value. The initial point used for iterative algorithm should be feasible for the optimization problem. We will discuss more details in the simulation part. 

\subsection{Update $(\lambda_k, \beta_k)$}
In this part, we give descriptions for updating the auxiliary variables $\lambda_k$ and $\beta_k$.

Notice that in \emph{Lemma 1}, the optimal solution $\tilde{p}_k^*, t_k^*,$ and $f_k^*$ should also satisfy the following system conditions:
\bea
\beta_k  E_k (\tilde{p}_k^*, t_k^*, f_k^*) - w_k R_k (\tilde{p}_k^*, t_k^*, f_k^*) = 0, \\
\lambda_k E_k (\tilde{p}_k^*, t_k^*, f_k^*) - w_k = 0.
\eea
Similarly, according to \cite{YJong}, we define functions for notational brevity. Specifically, let $T_j(\beta_j) = \beta_j E_k - w_k R_k$ and $T_{j+K} (\lambda_j) = \lambda_j E_k - 1$, $j \in \{1,2, \ldots, K\}$. The optimal solution for $\lambda_k$ and $\beta_k$ can be obtained by solving $\mathbf{T}(\lambda_k, \beta_k) = [T_1, T_2, \ldots, T_{2K}] = \mathbf{0}.$ We can apply iterative method to update the auxiliary variables. Specifically,
\bea
\lambda_k (i+1) \kern-1em&=&\kern-1em   (1-\theta (i)) \lambda_k (i) + \frac{\theta (i)}{E_k (\tilde{p}_k^*, t_k^*, f_k^*)}, \label{update1}\\
\beta_k (i + 1)\kern-1em &=& \kern-1em  (1-\theta(i)) \beta_k (i) + \theta (i) \frac{w_k R_k (\tilde{p}_k^*, t_k^*, f_k^*)}{E_k (\tilde{p}_k^*, t_k^*, f_k^*)} \label{update2} ,
\eea
where $\theta (i)$ is the largest $\theta$ that satisfies $||\mathbf{T} (\lambda_k (i) + \theta^l \mathbf{q}_{K+1:2K}^i, \beta_k(i) + \theta^l \mathbf{q}_{1:K}^i )|| \leq (1-z \theta^l) ||T(\lambda_k(i), \beta_k(i)||$,  $\mathbf{q}$ is the Jacobian matrix of $\mathbf{T}$, $l \in \{1,2, \ldots\}$, $\theta_l \in (0,1)$,  and $z \in (0,1)$. This update is also available in our prior paper \cite{meMEC}.
To summarize, we list the detailed algorithm in \textbf{Algorithm 1}.
\begin{algorithm}[]

\caption{Secure Computation Efficiency Maximization Algorithm}
\begin{algorithmic}[1]
\STATE  {\bf Initialization:} the algorithm accuracy indicator $u_1$ and $u_2$, set $i=0$, give initial values for $ v_k^{(i)}, \theta_k^{(i)}, t_k^{(i)}, N_k^{(i)}$, $\tau_k^{(i)}$, $\lambda_k^{(i)}$ and $\beta_k^{(i)}$. 
\WHILE {$||\mathbf{T}(\lambda_k, \beta_k)|| > u_1$}
\WHILE {$|\alpha_k (j+1) - \alpha_k(j)| > u_2$}
\STATE Solve for problem $\mathbf{P}_4$, obtain the intermediate optimal values $\{t_k\}, \{f_k\},   \{m_k\}, \{ \tilde{p}_k \}, \{\tau_k\},$ and $\{ N_k \}$.
\STATE Let $j = j+1$.
\ENDWHILE
\STATE Let $i = i+1$, update auxiliary variables $\lambda_k (i+1)$ and $\beta_k (i+1)$ from (\ref{update1}) and (\ref{update2}).
\ENDWHILE
\STATE Output the optimal computation efficiency.

\end{algorithmic}
\end{algorithm}

\section{Numeric Evaluation} 
In this section, simulation results from our proposed scheme and algorithm are presented.  Parameters for the simulation are given as follows. We assume the bandwidth for the system is $B = 200$ \emph{KHz}, number of users $K = 2$, time threshold $T = 1$ \emph{s}. For local computation, the CPU need $C_k = 1000$ operations to process one bit of data. In addition, the scaling factor for energy consumption $\epsilon_k = 1 \times 10^{-24}$, CPU for each user has a maximum frequency $f_k^\text{max} = 1 \times 10^9$ \emph{Hz}. For offloading, we assume the channel from user to the server to be $h_k^2/ \sigma^2 = H_k h_0$, where $h_0$ is the normal Gaussian variable.  Similarly, $g_k^2/ \sigma^2 = G_k h_0$. The value of $H_k$ and $G_k$ will be given later. We apply no bias to both users hence set $w_1 = w_2 = 1$. Lastly, the maximum allowed energy consumption is $E_k = 1$ \emph{Joule}. 

In Fig. \ref{1}, we show the convergence performance of our iterative algorithm. Here, we set $H_1 = 7, H_2 = 5$, $G_1 = G_2 = 1$, and $L_1 = L_2 = \{50000, 60000\}$. As a typical case, we only present the result for optimal time allocation $t_k$ here. It can be seen that our iterative algorithm has a good convergence speed, it only takes around 6 iterations to achieve the optimal value. In fact, we test with different initial points and manage to get the same performance. The other observation from Fig. \ref{1} is that, when the required computation bits becomes larger, the transmission time is also longer. Intuitively, this explains that more offloading bits are required.

\begin{figure}[h]
  \centering
  \includegraphics[width=3in]{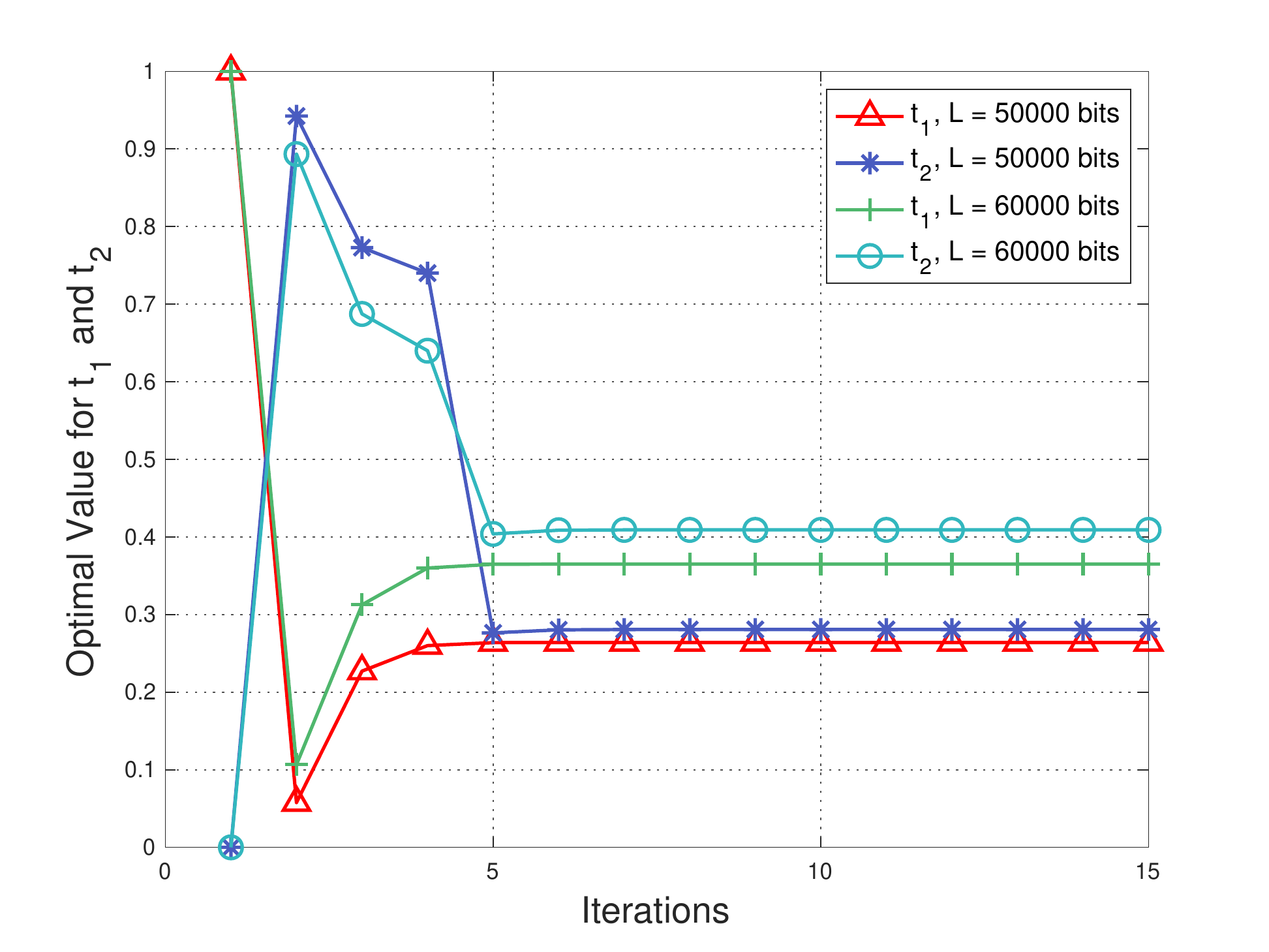}\\
  \caption{Iterative Algorithm Convergence Analysis}\label{1}
\end{figure}

\begin{figure}[h]
  \centering
  \includegraphics[width=3in]{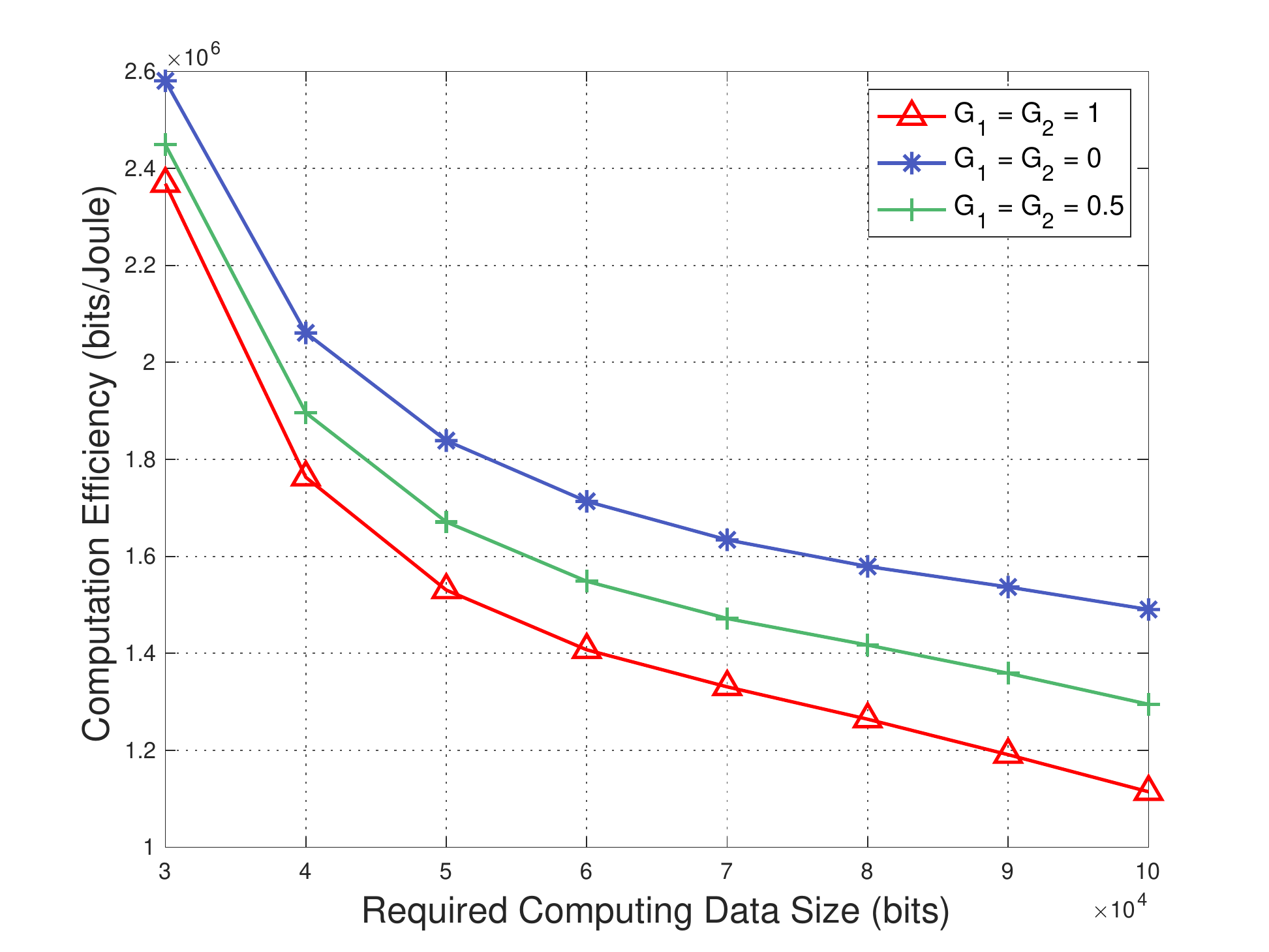}\\
  \caption{Computation Efficiency v.s. Required  Computation Bits under Different \emph{Eve} Channels }\label{2}
\end{figure}

In Fig. \ref{2}, the computation efficiency under different \emph{Eves} channels are presented. The x-axis is the number of required total computation bits. The first observation is that, under all scenarios, computation efficiency decreases with the increasing data size. If we break down the two parts for computation efficiency, we can easily see that pure local computing efficiency is square proportionally decreasing with the increasing of data size. Also, if the bit size is large, offloading part is also decreasing, due to, in part of the circuit power in the denominator. 

Additionally, Fig. \ref{2} also shows the relationship between computation efficiency with security threads from \emph{Eve}. Specifically, we set  channels from \emph{Eve} to users to be different. If the channel of \emph{Eve} is stronger, we see a setback in the performance, this is due to the impact from offloading, where achievable data rate is smaller.

\begin{figure}[h]
  \centering
  \includegraphics[width=3in]{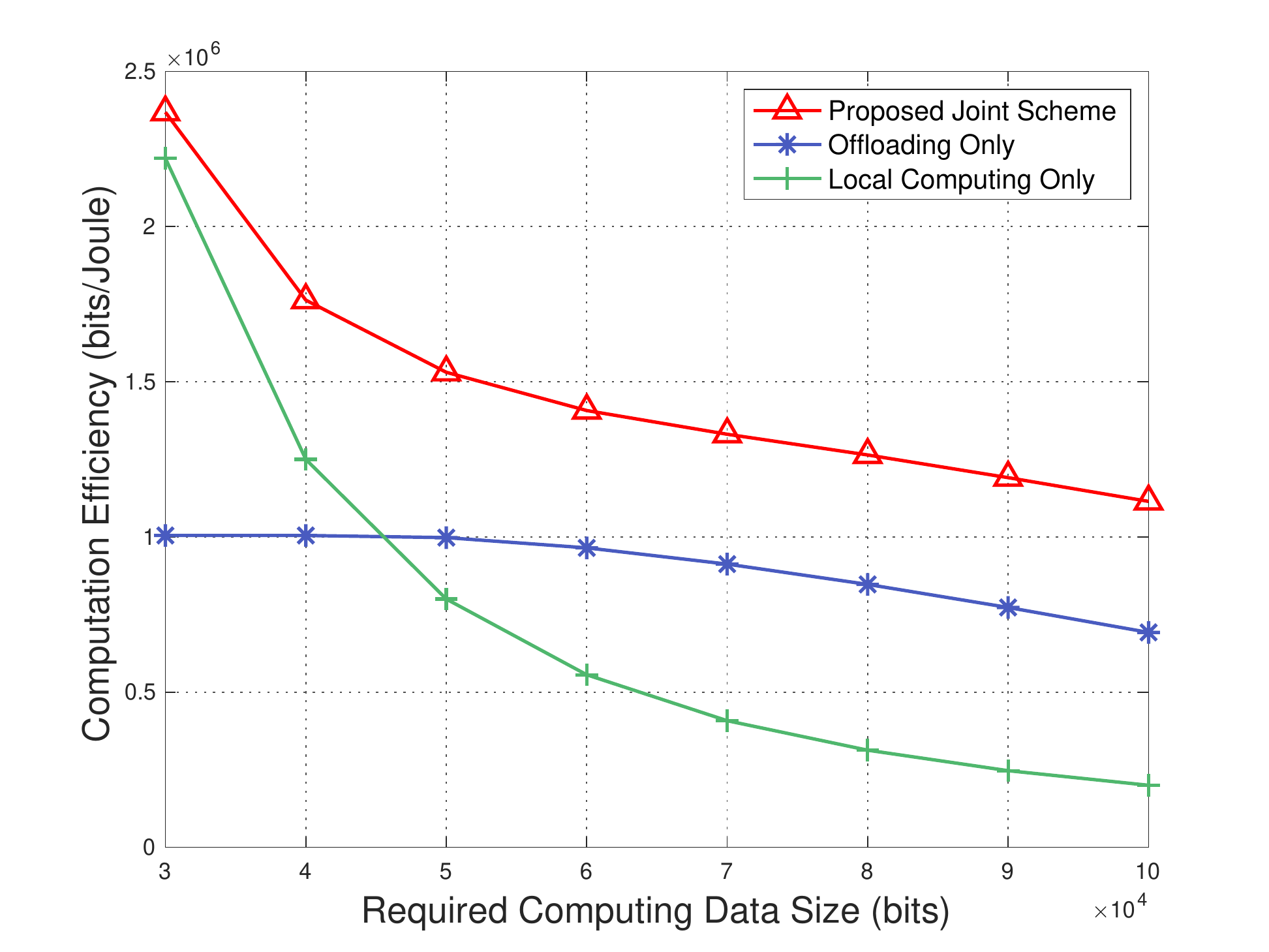}\\
  \caption{Computation Efficiency v.s. Required  Computation Bits under Different Computation Scheme}\label{3}
\end{figure}
Lastly, we compare our proposed joint offloading and local computation scheme with other two schemes: local computing only and offloading only. Here, we set $G_1 = G_2 = 1$. Clearly the proposed scheme outperforms both other two in terms of computation efficiency, which verifies the superiority of our scheme. Similar efficiency decreasing is also observed.

\section{Conclusions and Future Directions}
In this work, we studied the computation efficiency for a joint offloading and local computing scheme under possible $Eve$. We model the effect from $Eve$ with physical layer security and mutual entropy. An optimization problem is formulated which considers some practical constraints. This non-convex problem is transformed with SCA and a general ratio iterative algorithm. In the future, we plan to generalize the paper with multiple-antenna user/server and other access techniques. 

\end{document}